\documentclass{PoS}

\def\beq{\begin{equation}}
\def\eeq{\end{equation}}
\def\bea{\begin{eqnarray}}
\def\eea{\end{eqnarray}}
\def\beqn{\begin{eqnarray}} \def\eeqn{\end{eqnarray}}

\def\nn{\nonumber}
\def\Eq#1{Eq.~(\ref{#1})}

\newcommand\as{a_{\mathrm{S}}}
\newcommand\alphas{\alpha_{\mathrm{S}}}

\def\beq{\begin{equation}} \def\eeq{\end{equation}}
\def\beqn{\begin{eqnarray}} \def\eeqn{\end{eqnarray}}
 \def\to{\rightarrow}
\def\nn{\nonumber}

\title{
\vspace*{-2.5cm}
\begin{minipage}{\textwidth}
{\normalfont\small IFIC/16-79
\hspace{\fill} November 2016
}\\
\end{minipage}\\[60pt]
Mixed QCD-QED corrections to DGLAP equations}

\ShortTitle{Mixed QCD-QED corrections to DGLAP equations}

\author{\speaker{Germ\'an F. R. Sborlini}$^{\ a,b}$, Daniel de Florian$^c$ and Germ\'an Rodrigo$^a$\\\\
        $^a$Instituto de F\'{\i}sica Corpuscular, Universitat de Val\`{e}ncia -- 
Consejo Superior de Investigaciones Cient\'{\i}ficas, Parc Cient\'{\i}fic, E-46980 Paterna, Valencia, Spain.\\\
        $^b$Dipartimento di Fisica, Universit\`a di Milano and INFN Sezione di Milano,
I-20133 Milan, Italy.\\\
        $^c$International Center for Advanced Studies (ICAS), UNSAM, Campus Miguelete \\
25 de Mayo y Francia, (1650) Buenos Aires, Argentina.\\\\
        E-mail: \email{german.sborlini@ific.uv.es, deflo@unsam.edu.ar, german.rodrigo@csic.es}}

\abstract{We study the mixed effect of QCD and QED corrections to the evolution of parton distribution functions (PDFs). The Altarelli-Parisi splitting functions are extended to one order higher in QED, reaching ${\cal O}(\alpha \, \alpha_S^2)$ accuracy. This also involves extending DGLAP equations to include charge separation effects, that are ignored for pure QCD corrections. Besides that, these effects are crucial for the determination of the photon distribution, which plays an increasingly important role in nowadays phenomenological analysis.}

\FullConference{38th International Conference on High Energy Physics\\
                  3-10 August 2016\\
                 Chicago, USA}

\begin{document}

\section{Introduction}
The computation of higher-order corrections in QCD is a crucial part of the theoretical program in particle physics. In particular, the Altarelli-Parisi (AP) \cite{Altarelli:1977zs} splitting functions are a key component in these computations since they dictate the behaviour of scattering amplitudes in the collinear limit, besides establishing the evolution of parton distribution functions (PDFs) through the well-known DGLAP equations.

Due to the fact that ${\cal O}(\alphas^2)\approx {\cal O}(\alpha)$ for LHC processes, it becomes necessary to take into account also EW corrections. Thus, in this work, we present the computation of mixed QCD-QED corrections to the splittings functions and their effects in the evolution equations. The first noticeable modification is the presence of non-trivial photon and leptonic distributions. Thus, the evolution equations have to be generalized, which also implies new evolution kernels that appear at ${\cal O}(\alpha)$ and beyond. 

In order to exploit the previous knowledge of pure QCD corrections to AP kernels \cite{Curci:1980uw, Furmanski:1980cm,Ellis:1996nn}, we applied an
Abelianization algorithm that allowed to isolate the photon contributions and obtain the full QED correction by adjusting the colour and symmetry factors \cite{deFlorian:2015ujt,deFlorian:2016gvk}. These results complement our previous research on the QCD-QED splitting functions in the time-like regime \cite{Sborlini:2013jba,Sborlini:2014mpa,Sborlini:2014kla}, thus establishing a complete framework for obtaining higher-order mixed QCD-QED corrections in hadronic observables.

\section{Results and discussion}
The starting point consists in the extension of the usual DGLAP equations to deal with photon and lepton distributions. In the most general case, this leads to a coupled system of integro-differential equations. In order to simplify this problem, we change the PDFs basis as suggested in Ref. \cite{Roth:2004ti} and impose some physical constraints in the AP kernels. In particular, up to ${\cal O}(\alpha^2 \, \alphas^n)$, we have $P_{l q} \neq P_{q l}$ and $\Delta P_{fF}^S  \equiv 0$, where we use the notation $P_{ij} = \sum_{k,l} \, \as^k a^l \, P_{ij}^{(k,l)}$ with $\alpha \equiv 2\pi \, a$ and $\alphas \equiv 2\pi \, \as$, to perform the perturbative expansion of the splitting function associated to the process $j \to i + X$. The singlet and non-singlet distributions are properly defined in Refs. \cite{deFlorian:2015ujt,deFlorian:2016gvk}, as well as the remaining shorthand notation used across this article. After these considerations, the DGLAP coupled system is diagonalised partially, i.e.
\beqn
\nn && \frac{dq_{v_i}}{dt} =   P_{q_i}^-     \otimes q_{v_i}    \, , \quad \quad \frac{dl_{v_i}}{dt} =   P_{l}^-     \otimes l_{v_i}   \, , \quad \quad \frac{d \{\Delta_{2}^l,\Delta_{3}^l\}}{dt} =  P_{l}^+ \otimes \{\Delta_{2}^l,\Delta_{3}^l \}  \, ,
\\ && \frac{d \{ \Delta_{uc} , \Delta_{ct} \}}{dt} =  P_{u}^+ \otimes \{ \Delta_{uc} , \Delta_{ct}  \} \, , \quad \quad \frac{d \{ \Delta_{ds} ,\Delta_{sb}  \}}{dt} =  P_{d}^+ \otimes \{ \Delta_{ds} ,\Delta_{sb}  \} \, ,
\label{eq:evoluciondiagonal}
\eeqn
for the valence distributions $\{q_v,l_v\}$ and singlets $\{\Delta^l_2,\Delta^l_3,\Delta_{uc},\Delta_{ct},\Delta_{ds},\Delta_{sb}\}$. The evolution of the remaining components is dictated by more complicated equations \cite{deFlorian:2015ujt,deFlorian:2016gvk}, which we avoid here for the sake of simplicity.

\begin{figure}[ht]
\begin{center}
\includegraphics[width=0.48\textwidth]{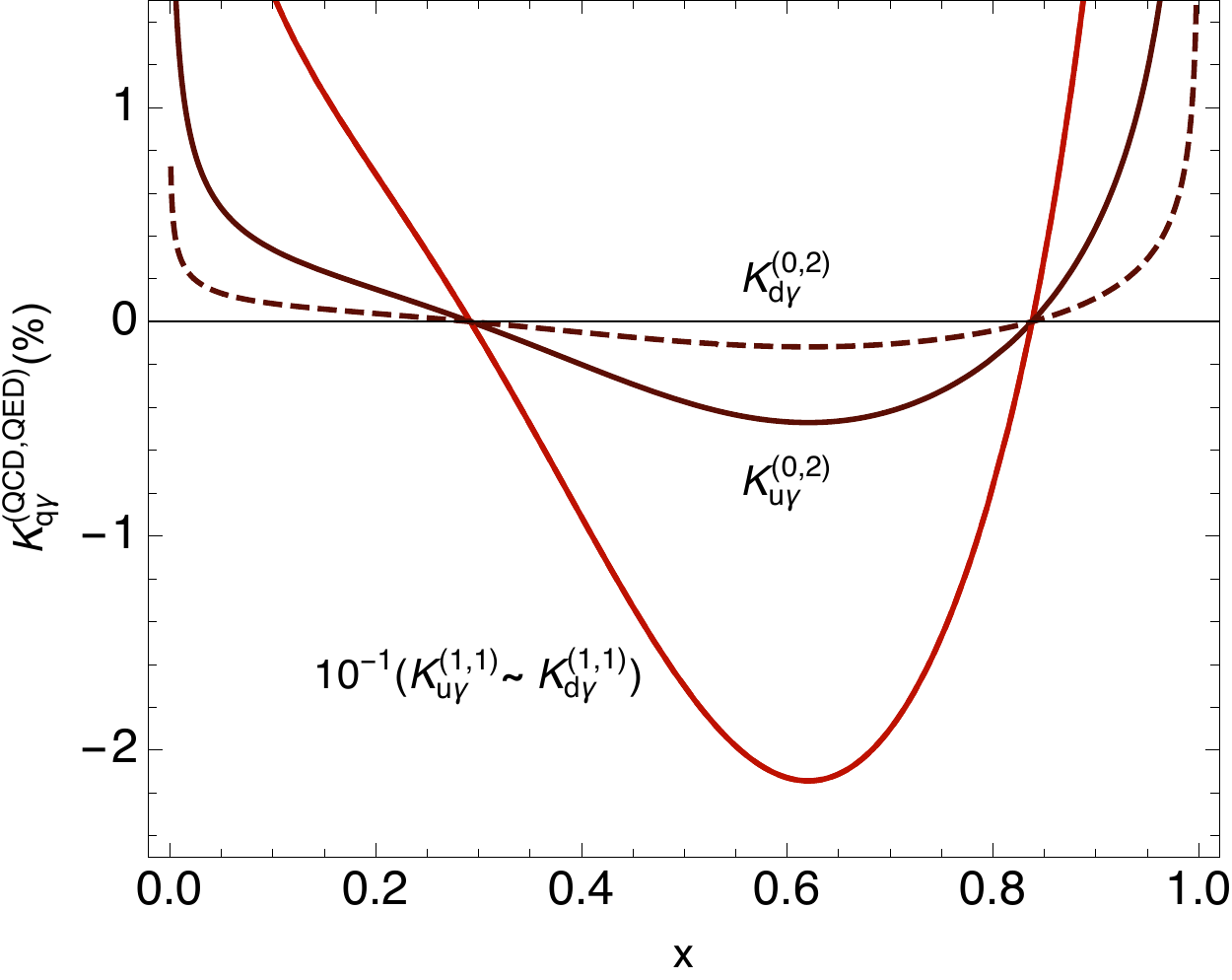} $\quad$
\includegraphics[width=0.48\textwidth]{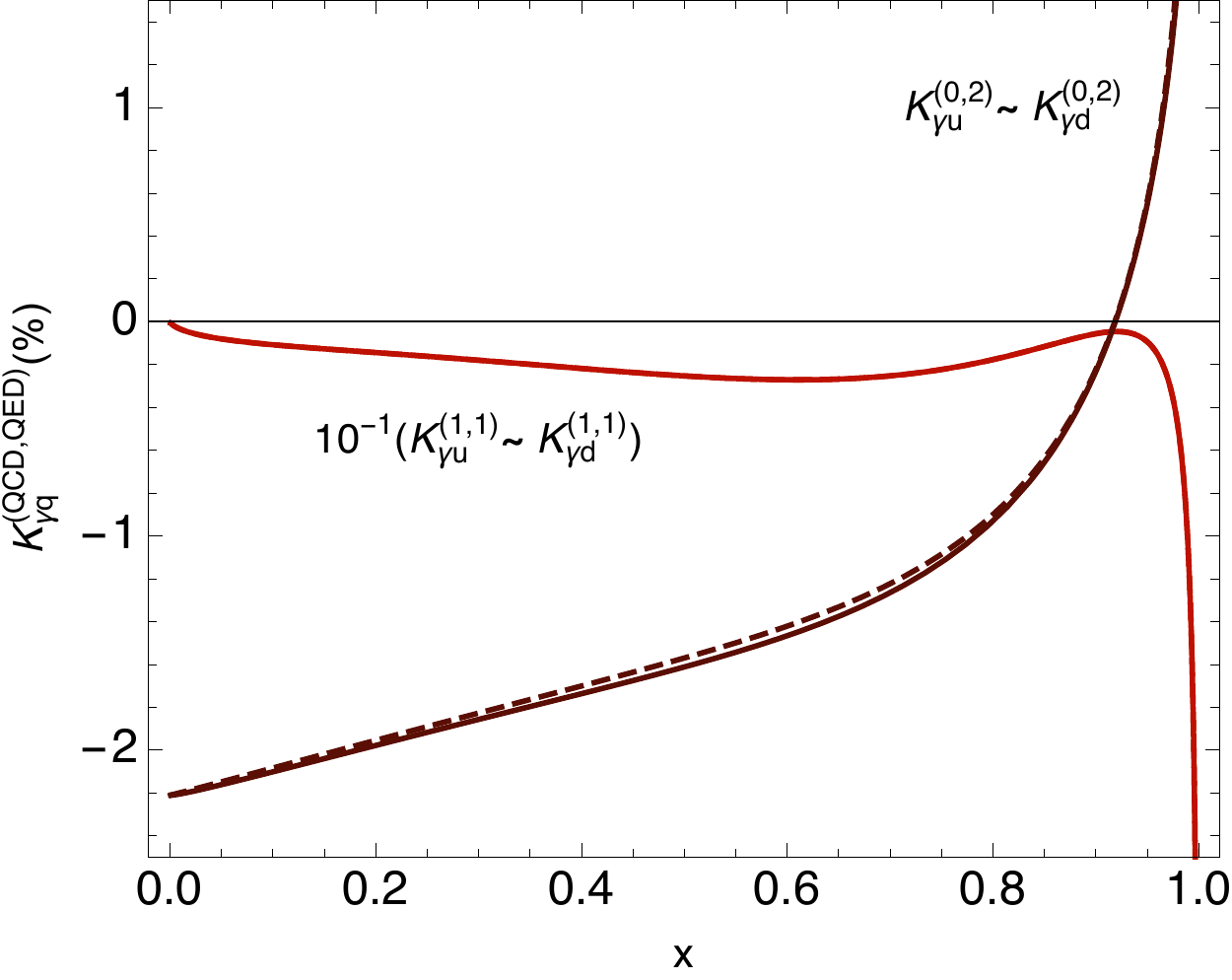} 
\caption{$K$ factors for the $q\gamma$ (left) and $\gamma q$ (right) splitting functions ($\%$). ${\cal O}(\alpha^2)$ and ${\cal O}(\alpha\,  \alphas)$ terms are included. An enhancement of the EM charge distinction in $P^{(i,j)}_{q \gamma}$ close to $x \approx 0.65$ is predicted.} 
\label{fig:QuarkPhoton}
\end{center}
\end{figure} 

On the other hand, we applied an Abelianization algorithm, as carefully described in Ref.~\cite{deFlorian:2015ujt}, to extract the Abelian contributions to the mixed QCD-QED kernels. Roughly speaking, the central idea is \emph{replacing gluons with photons}, but taking care of counting factors and vanishing topologies. The pure NLO QCD results (i.e. ${\cal O}(\alphas^2)$ contain diagrams with up to two gluons; thus, we can replace one gluon by a photon to recover ${\cal O}(\alpha \, \alphas)$ corrections, or replace both gluons to obtain the ${\cal O}(\alpha^2)$ kernels. It is worth appreciating that keeping track of the contributing topologies was crucial to check our results.

The evolution equations listed in \Eq{eq:evoluciondiagonal}, together with the non-diagonal ones associated to the remaining distributions, are subject to physical constraints in the end-point region (i.e. $x=1$). Explicitly, by imposing momentum and fermion number conservation for the proton, we have
\beq
0 = \frac{d P}{dt} = \int_0^1 dx \, x \, \left(\frac{dg}{dt}+\frac{d\gamma}{dt}+\sum_f \left(\frac{df}{dt}+\frac{d\bar{f}}{dt} \right) \right)  \, , \quad \quad \int_0^1 dx \, P^{-}_f = 0 \, ,
\eeq
which gives rise to a collection of equations that constrains the $\delta(1-x)$ term in $P_{gg}$, $P_{\gamma \gamma}$ and $P_{ff}$, for both leptons and quarks. The coefficients obtained from imposing the sum-rules agree with those that we get by applying the Abelianization algorithm to the pure QCD end-point coefficients. Again, this constitutes a second cross-check of our results.

In order to illustrate the effects of the computed corrections, we plot the quark-photon $K$-factors, i.e. $K^{(i,j)}_{ab} = \as^i \, a^j \, {P^{(i,j)}_{ab} (x)/P^{\rm LO}_{ab}(x)}$, where the LO is defined as $P^{\rm LO}_{ab} = \as \,  P^{(1,0)}_{ab} + a \, P^{(0,1)}_{ab}$. The results are shown in Fig. \ref{fig:QuarkPhoton}. We appreciate that, besides providing a non-negligible percent-level correction in comparison with the LO QED results, mixed QCD-QED contributions enhance charge separation at ${\cal O}(\alpha^2)$. This might originate important effects in the determination of photon distributions from global analysis.

\section{Conclusion}
We have obtained the complete set of Altarelli-Parisi kernels to ${\cal O}(\alpha \, \alpha_S)$ and ${\cal O}(\alpha^2)$, including those related to leptonic and photon densities. These new distributions mix in the evolution with the usual QCD parton distributions, as a consequence of the photon-mediated interaction starting at two-loops in QED.

For computing these kernels, we started from the well-known splitting functions at two-loops in pure QCD (i.e. ${\cal O}(\alphas^2)$). After a careful application of an Abelianization algorithm, we manage to identify and extract the corresponding photon contributions.

On the other hand, we studied the phenomenological consequences of these corrections to the evolution equations and splitting kernels. We confirmed that two-loop corrections are negligible for the pure quark kernels, but become sizeable in $P_{qg}$ and $P_{gq}$ for small values of the momentum fraction. In fact, the QED interactions could generate corrections of ${\cal O}( 2 \, \%)$ for photon initiated splittings. This was recently confirmed in Ref. \cite{Manohar:2016nzj}, where a complete analysis of the photon distribution was presented.

Finally, it is worth appreciating that the knowledge of the full set of AP kernels is crucial not only for improving the determination of the photon distribution, but also for achieving a fully consistent treatment of infrared (IR) singularities in the computation of hadronic observables with EW corrections. Recent works in EW corrections to gauge boson production \cite{Bonciani:2016wya} address this issue, and provide a practical application of higher-order AP splittings. From the theoretical point of view, the mixed QCD-QED splitting kernels are vital ingredients to built the counter-terms that cancel the IR divergences in initial-state radiation, thus allowing to safely compute up to ${\cal O}(\alpha^2)$ contributions.

\section*{Acknowledgments}
This work is partially supported by CONICET, ANPCyT, by the Spanish Government, by EU ERDF funds (grants FPA2014-53631-C2-1-P and SEV-2014-0398) and by GV (PROMETEU II/2013/007).

\end{document}